\documentclass[12pt]{article}
\usepackage{amssymb}  
\usepackage{amsthm}
\usepackage{amsmath}
\usepackage[left=1.5in,right=1.5in,bottom=1.5in,top=1.5in]{geometry}
\usepackage{graphicx}
\usepackage{authblk}
\usepackage[utf8]{inputenc}

\usepackage[backend=biber,style=nature,sorting=ynt]{biblatex}

\title{Mapping low-dimensional dynamics to high-dimensional neural activity: A derivation of the ring model from the neural engineering framework}
\author[1,2,*]{Omri Barak}
\author[3,*]{Sandro Romani}
\affil[1]{Rappaport Faculty of Medicine, Technion Israel Institute of Technology, Haifa 32000, Israel}
\affil[2]{Network Biology Research Laboratories, Technion - Israel Institute of Technology, Haifa 32000, Israel }
\affil[3]{Janelia Research Campus, Howard Hughes Medical Institute, Ashburn, Virginia 20147, USA}
\affil[*]{e-mail: omri.barak@gmail.com, sandro.romani@gmail.com}

\begin{document}
	\maketitle

\section{Abstract}

Empirical estimates of the dimensionality of neural population activity are often much lower than the population size. Similar phenomena are also observed in trained and designed neural network models. These experimental and computational results suggest that mapping low-dimensional dynamics to high-dimensional neural space is a common feature of cortical computation. Despite the ubiquity of this observation, the constraints arising from such mapping are poorly understood. Here we consider a specific example of mapping low-dimensional dynamics to high-dimensional neural activity -- the neural engineering framework. We analytically solve the framework for the classic ring model -- a neural network encoding a static or dynamic angular variable. Our results provide a complete characterization of the success and failure modes for this model. Based on similarities between this and other frameworks, we speculate that these results could apply to more general scenarios.

\section{Introduction}

The activity of large neuronal populations can be described, a priori, in a space whose dimension is comparable to the size of the population. Nonetheless, large-scale neuronal recordings consistently exhibit low-dimensional activity dynamics \cite{gao_simplicity_2015}. Such low dimensionality could arise if these large populations are actually representing low-dimensional features of the external world, or if the network dynamics is somehow constrained to be low dimensional. 

Recent studies on artificial networks that were trained to solve cognitive tasks also revealed similar low-dimensional structures \cite{sussillo_opening_2013,mante_context-dependent_2013,russo_motor_2018,beer_dynamics_2018}. In this case, neither the connectivity structure of the network, nor its desired dynamics were specified ahead of time. The analyses performed on these networks showed that this low dimensionality emerged to perform a given task.

Classic works on task-performing neural networks focused on designing the connectivity, rather than training it. These designs often relied upon a low-dimensional structure of connectivity and of activity \cite{ben-yishai_theory_1995,hopfield_neural_1982}. This connectivity was inspired by insightful observations, rather than relying on  a systematic or algorithmic procedure.

Both the experimental and computational results suggest that mapping low-dimensional dynamics to high-dimensional neural space is a common feature of cortical computation \cite{gao_simplicity_2015}. Despite all this progress, most of the results remain empirical, lacking a theoretical framework that exposes constraints arising from such mapping (But see \cite{mastrogiuseppe_linking_2018,rivkind_local_2017}).

There are algorithmic frameworks that explicitly highlight the mapping from low to high dimensionality. This class of models allows to specify a desired low-dimensional dynamics, and obtain connectivity implementing these dynamics in high-dimensional activity space \cite{eliasmith_neural_2004, boerlin_balanced_2012,thalmeier_learning_2016}. These models can be considered as an intermediate between trained and designed networks, and as such natural candidates for an in-depth study of this mapping. We focus on one concrete proposal for a mapping - Neural Engineering Framework (NEF, \cite{eliasmith_neural_2004}) - and solve it analytically for a specific task \cite{ben-yishai_theory_1995}. 
We show conditions for success and failure of such models, and analyze the underlying causes.

\newpage

\section{General framework}

To describe an equivalence between dynamical systems in different spaces, we need a mapping between these spaces. To this aim, we define a mapping from a low $d$-dimensional
feature vector $x\in R^{d}$ to a high $N$-dimensional vector of firing rates $\hat{r} \in R^{N}$, $\hat{r}(x)=\mathcal{F}(x)$ (Figure \ref{fig:framework}A). In a similar manner, we define a mapping from rate vectors $r$ to feature vectors $\hat{x}\in R^d$, $\hat{x}=\mathcal{G}(r)$.

To render the two mappings consistent with each other, we choose $\mathcal{F}$ and $\mathcal{G}$ such that (Figure \ref{fig:framework}B) 

\begin{equation}
x = \hat{x}(\hat{r}(x))
\label{eq:general_equiv}
\end{equation}

\noindent for relevant $x$ values. Having defined these mappings, we turn to dynamics (Figure \ref{fig:framework}C). We would like to construct a high-dimensional neural network that implements a desired low-dimensional dynamical system in feature space. In more formal terms, the two dynamical systems are:
\begin{eqnarray}
\dot{x}&=&h(x) \\
\dot{r}&=&-r+f(Jr)
\label{eq:rate_model}
\end{eqnarray}

\noindent where $J\in \mathcal{R}^{N\times N}$ is the network connectivity, $f$ is a static nonlinearity applied element-wise representing the input-to-rate transformation of the neuron. We would like to choose the connectivity $J$ such that projections of $r$ trajectories will be equal to $x$ trajectories. If $r(0) = \hat{r}(x(0))$, then for $t>0$:
\begin{eqnarray}
\hat{x}(r(t))&=&x(t) 
\label{eq:general_identity_dyn}
\end{eqnarray}

For concreteness, we will now analyze a specific example of such an equivalence - the neural engineering framework \cite{eliasmith_neural_2004}.

\section{Neural Engineering Framework}

\begin{figure*}
		\centering
		\includegraphics[width=0.95\linewidth]{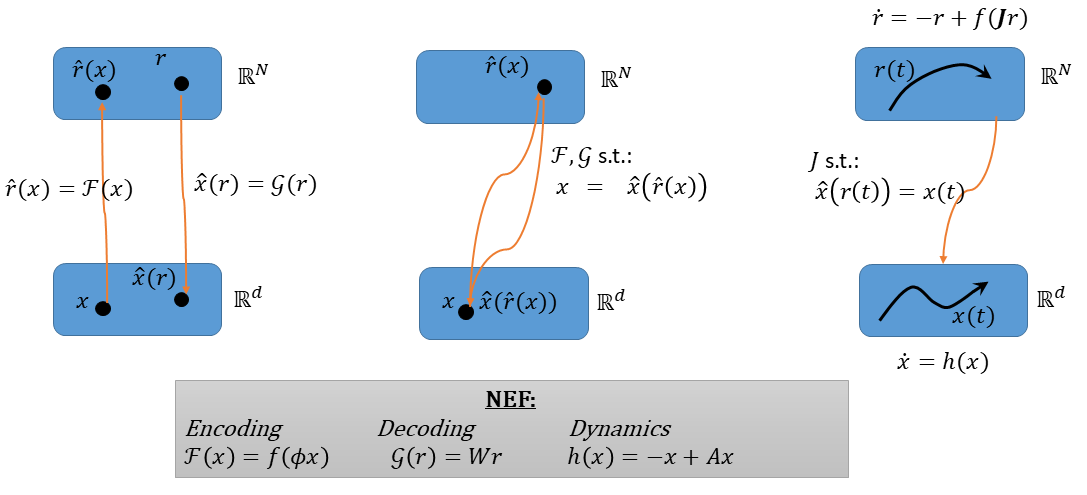}		
		\caption{\textbf{Mapping between low-dimensional feature dynamics and high-dimensional neural activity space}. \textbf{(A)} A low dimensional
feature vector $x\in R^{d}$ is encoded by a high dimensional rate vector $\hat{r} \in R^{N}$, $\hat{r}(x)=\mathcal{F}(x)$, where $N \gg d$. Conversely, a decoder $\hat{x}=\mathcal{G}(r)$ maps rate to feature vectors. \textbf{(B)} Imposing consistency between the mappings. \textbf{(C)} The dynamics of the firing rate vector $r$ arise from a network with connectivity $J$. The objective is to find a connectivity $J$ such that the decoded feature dynamics will match a desired dynamics $\dot{x}=h(x)$. \textbf{(D)} The Neural Engineering Framework (NEF) uses the same static nonlinearity $f$ for the neural network dynamics and for encoding. Decoding is done via a linear readout, and feature dynamics are linear. }
		\label{fig:framework}
	\end{figure*}

The neural engineering framework \cite{eliasmith_neural_2004} discusses how to implement a low-dimensional dynamical system with a network of rate or spiking neurons. For simplicity, we will consider a representative form of this framework (Figure \ref{fig:framework}D).
The mapping $\mathcal{F}$ is given by a nonlinear function of a matrix $\phi\in R^{N\times d}$ operating on the features. The nonlinearity is chosen to be identical to the nonlinearity of the neurons in the network 

\begin{equation}
\hat{r}(x) = \mathcal{F}(x)=f(\phi x)
\end{equation}.

The matrix $\phi$ represents the mapping from features to neuronal inputs, and therefore $f(\phi x)$ is akin to a tuning curve.  For the mapping $\mathcal{G}$, since $d \ll N$, we use a linear decoder 
\begin{equation}
\hat{x}(r) = \mathcal{G}(r)=Wr
\end{equation}.

With these choices Equation \ref{eq:general_equiv} becomes
\begin{equation}
x=Wf(\phi x)
\label{eq:identity}
\end{equation}

\noindent for all $x$ values of interest. Note that in the original description \cite{eliasmith_neural_2004},
$W$ was defined as a least square solution, but we are interested
in analytically tractable cases where an exact equality is possible.

 For simplicity we consider the desired low-dimensional dynamics as linear (the nonlinear case is described in Appendix A):
\begin{equation}
\dot{x} = -x+Ax
\label{eq:feature_dynamics}
\end{equation}

The evolution of the projection $\hat{x}=Wr$ can be obtained from the $r$-dynamics (Equation \ref{eq:rate_model}):

\begin{equation}
    \dot{\hat{x}} = -\hat{x} + Wf(Jr)
\end{equation}

Our objective is to achieve the target dynamics $\dot{\hat{x}} = -\hat{x} + A\hat{x}$. Applying Equation \ref{eq:identity}, we can rewrite our objective as:

\begin{equation}
\frac{d\hat{x}}{dt} = -\hat{x} + Wf(\phi A W r)
\end{equation}

Thus, by defining 
\begin{equation}
J=\phi AW,
\label{eq:J}
\end{equation}
we obtain the desired dynamics for $\hat{x}$, which is induced by the network dynamics $\dot{r}=-r+f(Jr)$. To derive Equation \ref{eq:J} we implicitly defined the $x$ values of interest (and thus $W$) in equality \ref{eq:identity} as $Ax$, resulting in:

\begin{equation}
    Ax=Wf(\phi Ax)
    \label{eq:identity2}
\end{equation}

\noindent where we consider all $x$ values that appear in the dynamics described by Equation \ref{eq:feature_dynamics}. This condition is important, because in general it is not possible to have a consistent mapping for all $x$ values when using linear mappings to approximate nonlinear ones. The consequences of different choices are explored in Appendix E.

\section{The ring model}

The linear decoder $W$ is usually estimated numerically, but here we
wish to study an example where it can be derived analytically. For
that we consider the well studied ring model \cite{ben-yishai_theory_1995}, where a periodic variable
$\psi$ is represented by neurons labeled by their preferred angle
$\theta\in[-\pi,\pi]$. The model is described by the following rate dynamics:


\begin{eqnarray}
\dot{r}_\theta&=&-r_\theta+\sigma(Jr_\theta) \\
J_{\theta,\theta'}&=&J_{0}+J_{1}\cos(\theta-\theta') 
\label{eq:ring}\\
\sigma(z)&=&[z+I_e]_{+}
\label{eq:ring_transfer}
\end{eqnarray}

\noindent with $[z]_{+}=\text{max}(z,0)$, and the external drive $I_e>0$ which is required to obtain non-zero solutions \cite{ben-yishai_theory_1995}. It is known that for particular combinations
of $J_{0},J_{1}$ there exists a marginally stable "bump" solution where the
activity of the neurons is given by

\begin{eqnarray}
r_{\theta}\propto[\cos(\theta-\psi)-\cos(\theta_{C})]_{+}
\label{eq:ring_tuning}
\end{eqnarray}

\noindent where $\cos \theta_C$, which determines the width of the activity bump, is a function of $J_1$ (Figure 2, marginal phase), defined by the relationship:

\begin{eqnarray}
J_{1}=g_{1}^{-1}(\theta_C)
\label{eq:ring_J1}
\end{eqnarray}
where
\begin{eqnarray}
g_{1}(\theta_C)=\int\frac{d\theta}{2\pi}[\cos\theta-\cos\theta_C]_+\cos\theta
\label{eq:ring_g1}
\end{eqnarray}

	\begin{figure*}
		\centering
		\includegraphics[width=0.95\linewidth]{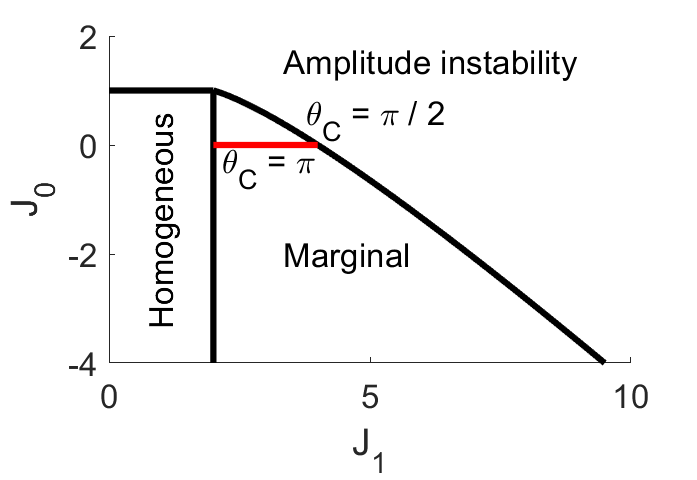}
		\caption{ \textbf{Ring model parameters attainable by NEF}. The ring model is characterized by a two-parameter ($J_0,J_1$) family of network connectivities. These parameters define regions of qualitatively different solutions. Activity decays in the homogeneous regime, diverges in the instability regime, and forms a stable "bump" in the marginal regime. The NEF recovers the ring model, but with the constraints of $J_0=0$ and $2<J_1<4$ (red line).}
		\label{fig:Phase diagram}
	\end{figure*}

\section{Applying the framework to the ring model}

The NEF links encoding and dynamics by requiring that the same nonlinearity $f$ is used both in the definition of the mapping from the low-dimensional to the high-dimensional space, and in the dynamics of the neural network. 

We thus define the nonlinearity $f(z)=[z+I_e]_+$ exactly as in Equation \ref{eq:ring_transfer}, and proceed to check whether we can recover the tuning curves described by Equation \ref{eq:ring_tuning}.

We are interested in the marginal phase, which has a stationary bump solution in the periodic feature space. We thus define $x\in R^{2}$
to be 

\begin{equation}
x=\left(\begin{array}{c}
\cos\psi\\
\sin\psi
\end{array}\right)
\end{equation}

As in the ring model, we parameterize the neurons by an index $\theta$,
and notice that if we define $\phi \in R^{N \times 2}$ as

\begin{equation}
\phi=\left(\begin{array}{cc}
\cos\theta & \sin\theta\end{array}\right)
\end{equation}

\noindent we obtain tuning curves similar to Equation \ref{eq:ring_tuning}:

\begin{equation}
r=[\cos(\theta-\psi)+I_e]_{+}
\end{equation}

In order for this to exactly match Equation \ref{eq:ring_tuning}, we need the external drive to be $I_e = -\cos \theta_C$. For narrow bumps, $\theta_C \in [0,\pi/2]$, we obtain $I_e<0$ in violation of the ring model requirements. We return to this topic later.


The ring model has a stationary bump, corresponding to $\dot{\psi}=0$ , implying $A=I$ in Equation \ref{eq:feature_dynamics}. 
It is now possible to find $W$ that will fulfill equation \ref{eq:identity2}. In
our case the solution is described by two equations:

\begin{equation}
\cos\psi=N^{-1}\sum W_{1,\theta}[\cos(\theta-\psi)-\cos\theta_C]_{+}
\end{equation}

\begin{equation}
\sin\psi=N^{-1}\sum W_{2,\theta}[\cos(\theta-\psi)-\cos\theta_C]_{+}
\end{equation}

In the limit of $N \rightarrow  \infty$, we can replace the sum with an integral, and replace the two equations with a single one for a complex variable  $w(\theta)=W_{1,\theta}+iW_{2,\theta}$:

\begin{equation}
e^{i\psi}=\int\frac{d\theta}{2\pi}w(\theta)g(\theta-\psi)
\label{eq:convolution}
\end{equation}

\noindent where $g(z)=f(\cos z)$. Equation \ref{eq:convolution} is a convolution, and hence lends
itself to a solution in Fourier space:

\begin{equation}
\delta_{k,1}=w_{k}\tilde{g}_{k}
\end{equation}

In general, we expect $\tilde{g}_{k}\neq0$ for all $k$. However, choosing $W$ with minimal norm would lead to:

\begin{equation}
w_{k}=\delta_{k,1}\tilde{g}_{k}^{-1}
\end{equation}

\noindent which, when transformed back from Fourier space results in:

\begin{equation}
W=\frac{1}{\tilde{g}_{1}}\left(\begin{array}{c}
\cos\theta\\
\sin\theta
\end{array}\right)
\label{eq:W_and_g1}
\end{equation}

\noindent where $\tilde{g}_{1}(\theta_C)=\int\frac{d\theta}{2\pi}[\cos\theta-\cos\theta_C]_+\cos\theta$. Note this is exactly the same $g_1$ of the ring model (Equation \ref{eq:ring_g1}). Finally, using Equation \ref{eq:J}, the resulting connectivity is

\begin{equation}
J_{\theta,\theta'}=g_{1}^{-1}(\theta_C)\cos(\theta-\theta')
\end{equation}

We thus recover the original ring model (Equation \ref{eq:ring}), but with the parameter
$J_{0}=0$  determined by this procedure. The stability of the ring model is known for all values of $J_{0}$ and
$J_{1}$. In particular, for $J_{0}=0$, a stable bump exists (marginally
stable regime, solid red line in Figure \ref{fig:Phase diagram}) for $g_{1}^{-1}(\pi) < J_{1} < g_{1}^{-1}(\pi/2)$. Since $\theta_C$ is a function of $J_1$ (Equation \ref{eq:ring_J1}), having narrow bumps requires $J_1>4$. This is also consistent with the definition of $I_e$ above. We now see that the solution provided by NEF prohibits the framework from generating a ring attractor with narrowly tuned neurons.

\section{Mapping between dynamical systems}

In the neural engineering framework, tuning curves for each neuron obey  the relationship $r=f(\phi x)$. If the activity of the neurons always maintains this relationship, then a trajectory $x(t)\in R^d$ should correspond to a $d$-dimensional manifold	 $r(t)=f(\phi x(t))$. To assess whether this holds, we assume a slight deviation from this manifold at a certain time $t$ (Figure \ref{fig:deviation}):

	\begin{figure*}
		\centering
		\includegraphics[width=0.95\linewidth]{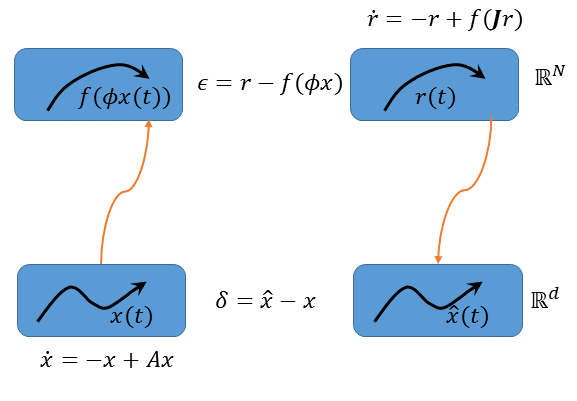}		
		\caption{ \textbf{Mapping between dynamical systems.} A desired low-dimensional trajectory $x(t)$ is mapped to a manifold $f(\phi x(t))$, but the high-dimensional dynamics $r(t)$ might deviate from this manifold.  }
		\label{fig:deviation}
	\end{figure*}

\begin{equation}
r(t)=f(\phi x(t))+\epsilon(t)
\end{equation}

We then follow the dynamics of this deviation by considering the simultaneous evolution of the low-dimensional and high-dimensional dynamics:

\begin{eqnarray}
\frac{dx}{dt} &=& -x + Ax \\
\frac{dr}{dt} &=& -r + f( Jr ) \\
\frac{d\epsilon}{dt} &=& \frac{dr}{dt}-f'(\phi x)\phi \frac{dx}{dt}
\end{eqnarray}

For small $\epsilon$, we arrive at:

\begin{eqnarray}
\frac{dr}{dt} &=& -r + f( Jf(\phi x) ) + f'(Jf(\phi x))J \epsilon \\
\frac{d\epsilon}{dt}&=&\left[-1+f'\left(J f(\phi x)\right)J\right]\epsilon+\left[ f\left(J f(\phi x\right))-f(\phi x)-f'(\phi x)\phi\frac{dx}{dt} \right]
\label{eq:deviation}
\end{eqnarray}

The dynamics of the deviation from the manifold is given by two terms
-- one that depends on $\epsilon$, and one that doesn't. If the second term is nonzero, then the manifold cannot be stable. There are two special cases in which this term vanishes.  For the case
of a fixed point dynamics $dx/dt=0$, we have $A=I$ and thus the identity \ref{eq:identity2} becomes
$x=Wf(\phi x)$. Since $J=\phi A W$, this causes the second term to vanish,
and stability is given by the first one. The other case is a linear $f$ (we assume $f(x)=x$ without loss of generality), for which the identity \ref{eq:identity2} implies $W\phi=I$. In this case, even for $dx/dt \neq 0$, the second term vanishes. Since the second term only vanishes for a linear $f$ or fixed point dynamics, we expect deviations from the manifold for almost all cases. 

\section{Applying the framework to the dynamic ring model}

The above arguments were not specific for the ring model. We now examine their implications in that specific case. First, the Jacobian is given by:

\begin{equation}
-\delta_{\theta,\theta'}+f'(\phi x)J
\end{equation}

We can calculate the Jacobian for our choice of $\phi$ and $f$,
around a point $x$ characterized by $\psi=0$ :

\begin{equation}
-\delta_{\theta,\theta'}+\frac{1}{g_1}H\left(\cos\theta-\cos\theta_{C}\right)\cos(\theta-\theta')
\label{eq:Jacobian}
\end{equation}

\noindent where $H$ is the heaviside function. The eigenvalues are (see Appendix B):

\begin{eqnarray}
\lambda_{1} &=& 0 \\
\lambda_2 &=& -1 + \frac{\theta_C+1/2 \sin2\theta_C}{\theta_C-1/2 \sin2\theta_C} \\
\lambda_k &=& 0 \text{ for }k>2
\end{eqnarray}

We see that indeed for wide tuning curves ($\theta_C>\pi/2$) we have $\lambda_2<0$ leading to marginal stability of the manifold. In this case, this is just a restatement of the stability of the ring model, and the eigenvalues are identical to those of the $J_0=0$ case \cite{hansel_methods_1998}.

To assess the contribution of the second term of Equation \ref{eq:deviation}, we expand the model to include a bump moving with a velocity $v$, giving rise to the following
dynamics:

\begin{equation}
\frac{dx}{dt}=-x+\left(\begin{array}{cc}
1 & v\\
-v & 1
\end{array}\right)x
\end{equation}

With this modification, equation \ref{eq:convolution} becomes:

\begin{equation}
(1-i v)e^{i\psi}=\int\frac{d\theta}{2\pi}w(\theta)h(\theta-\psi)
\end{equation}

\noindent with

\begin{equation}
    h(z) = f( \cos z + v \sin z )
\end{equation}
leading to (see Appendix D)

\begin{equation}
J_{\theta,\theta'} \propto \left[\cos(\theta-\theta')+v\sin(\theta-\theta')\right]
\end{equation}

\begin{figure*}
		\centering
		\includegraphics[width=0.95\linewidth]{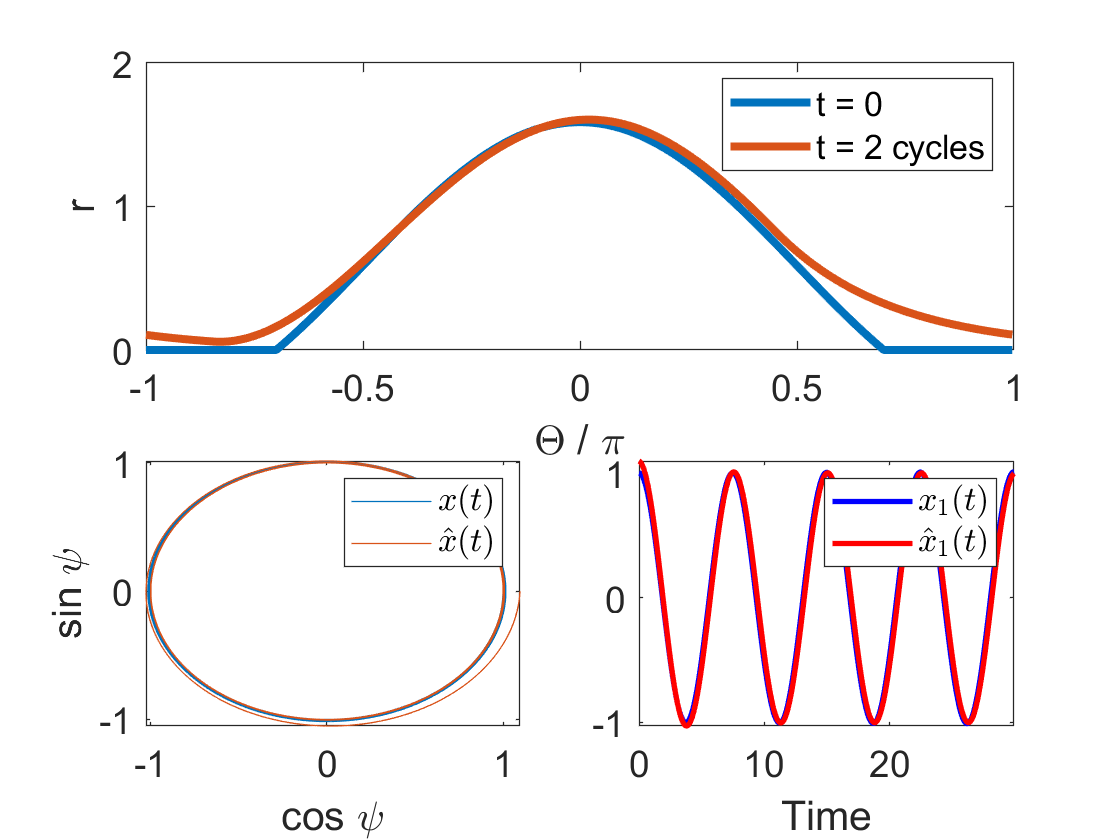}		
		\caption{\textbf{Deviation from desired dynamics.} Implementing a moving bump with the NEF gives rise to a deviation in the high-dimensional space. \textbf{A} Snapshots of population activity that starts on the desired manifold (blue), and eventually converges to a different profile (red). \textbf{B,C} Desired low-D trajectory (blue) vs. the actual one (red), showing no deviation in the low-dimensional dynamics.}
		\label{fig:moving}
	\end{figure*}

As noted above, the second term in equation \ref{eq:deviation} will vanish for a linear $f$, but is not expected to vanish in general. Figure \ref{fig:moving} shows a simulation of a wide bump with nonzero velocity that indeed deviates from the manifold. The initial condition is on the manifold (blue curve, Figure \ref{fig:moving}A), but with time the deviation from the manifold, $\epsilon$, grows and the activity of the neurons cannot be written as $r=f(\phi x)$ for any $x$ (red curve, Figure \ref{fig:moving}A). By construction, the framework guarantees that the desired $x$ dynamics are still obtained, as shown in Figure \ref{fig:moving}B,C. 

\section{Deformation of the manifold}

The example above illustrated the instability of the high-dimensional manifold. Despite this instability, the activity converged to a stable limit cycle. This suggests that perhaps the target manifold should be defined in a different manner. Indeed, in the original ring model with a moving bump, the same asymmetric tuning profile arises \cite{hansel_methods_1998}. This is because the input to a neuron is still cosine shaped, but it is passed through a nonlinearity and then low-pass filtered by the dynamics.

We thus define a new desired manifold:

\begin{equation}
    \frac{d\bar{r}}{dt} = -\bar{r} + f(\phi A x )
\end{equation}

and the deviation from it:
\begin{eqnarray}
    \bar{\epsilon} &=& r - \bar{r} \\
    \frac{d\bar{\epsilon}}{dt} &=& -\bar{\epsilon} + f(Jr) - f( \phi A x) \\
     &=& -\bar{\epsilon} + f(\phi A \hat{x}) - f( \phi A x) \\
     &=& -\bar{\epsilon} + f'( \phi A x) \phi A \delta
\end{eqnarray}

\noindent where $\delta=\hat{x}-x$ is the deviation of the low-D dynamics. Since, by the NEF construction, $\hat{x}$ and $x$ follow the same dynamics, $\delta$ can only deviate by a phase shift along the bump trajectory. This implies that $\bar{\epsilon}$ decays to zero, and the high-D dynamics are stable to perturbations around $\bar{r}$. We thus see that the dynamics modifies the tuning curves of neurons, but in a predictable manner.


\section{Discussion}

We provide the first  solvable example of the neural engineering
framework (NEF). This framework allows to specify desired tuning curves of the neurons to a feature of interest. NEF then provides a numerical solution for the connectivity between neurons that will implement the desired feature dynamics by the neural network. We show analytically NEF's ability to recover the well
known ring model, which describes how a periodic feature can be encoded by a recurrent neural network. We then show NEF's limitations when the feature is either static or dynamic.
The synaptic connectivity arising from NEF is identical to that of the ring model, but limited to a subset of parameter values. This subset translates to a limitation on the type of tuning curves that can be implemented by NEF -- narrow tuning curves are unstable. 

Introducing dynamics in the low-D feature gives rise to a second form of instability. Namely, even if the provided tuning curves can be stably maintained for a static feature, the shape of the tuning curves changes as the feature becomes dynamic. We show, however, how the modified tuning curves can be predicted.

It should be stressed that we study a specific variant of NEF. The full framework is much richer, and has been shown to successfully implement a wide range of challenging dynamics \cite{eliasmith_large-scale_2012,eliasmith_how_2013}. Additional aspects that we neglected in our derivation, such as noise injections, might rescue the instabilities we describe here. 

For instance, the limitation to wide tuning curves resulted from a lack of global inhibition. Obtaining global inhibition through the NEF would require artificially increasing the dimensionality of the represented feature (Appendix C). However, this solution is only apparent after analyzing the instability arising from the specific mapping under consideration.

Nevertheless, our study sheds light on properties of the general problem of mapping dynamical systems in this framework. Specifically, the instability of the manifold in the case of dynamic features was derived for the general case, and not just for the ring model. This indicates that tuning curve limitations and discrepancies may be a more general phenomenon.



Beyond the neural engineering framework, our results could apply to other scenarios. The  neural engineering approach
was recently adapted by directly using spike timing for both linear \cite{boerlin_predictive_2013} and nonlinear \cite{thalmeier_learning_2016} systems. Similar mappings of dynamical systems were also used in different contexts \cite{depasquale_using_2016}. It
would be interesting to study whether the same qualitative instabilities are observed in these frameworks as well.

It is also interesting to note the analogies between the connectivity arising from NEF ($J=\phi A W$) and that arising by other means in real or artificial circuits. The NEF low-rank connectivity can be interpreted as \textit{linearly} mapping between low- and high- dimensional dynamical systems, with the surprising result that this transformation also works for nonlinear neural dynamics. Recently, low-rank perturbations to connectivity have received attention \cite{mastrogiuseppe_linking_2018,rivkind_local_2017,schuessler_dynamics_2020,logiaco_model_2019}, both due to specific training protocols \cite{sussillo_generating_2009}, and due to observations of low dimensional neural activity in data \cite{gao_simplicity_2015}. Together, this implies that something akin to the NEF formula might arise through the unconstrained training of neural networks.


\section{Acknowledgments}

We thank Amichai Labin for helping with an initial version of this project. We thank Chen Beer and Ran Darshan for comments on the manuscript. We thank the Ostojic lab for fruitful discussions. OB is supported by the Israeli Science Foundation (346/16). SR is supported by the Howard Hughes Medical Institute. We thank Misha Tsodyks, in whose lab this project started 12 years ago.

\printbibliography

@book{hansel_methods_1998,
	title = {Methods in {Neuronal} {Modeling}. {From} {Synapse} to {Networks}. {Koch} {C} and {Segev} {I}, editors},
	publisher = {MIT Press, Cambridge, MA. Chapter Modeling Feature Selectivity in Local …},
	author = {Hansel, D. and Sompolinsky, H.},
	year = {1998}
}

@article{schuessler_dynamics_2020,
	title = {Dynamics of random recurrent networks with correlated low-rank structure},
	volume = {2},
	url = {https://link.aps.org/doi/10.1103/PhysRevResearch.2.013111},
	doi = {10.1103/PhysRevResearch.2.013111},
	number = {1},
	urldate = {2020-02-03},
	journal = {Physical Review Research},
	author = {Schuessler, Friedrich and Dubreuil, Alexis and Mastrogiuseppe, Francesca and Ostojic, Srdjan and Barak, Omri},
	month = feb,
	year = {2020},
	pages = {013111}
}

@article{logiaco_model_2019,
	title = {A model of flexible motor sequencing through thalamic control of cortical dynamics},
	copyright = {© 2019, Posted by Cold Spring Harbor Laboratory. This pre-print is available under a Creative Commons License (Attribution-NonCommercial-NoDerivs 4.0 International), CC BY-NC-ND 4.0, as described at http://creativecommons.org/licenses/by-nc-nd/4.0/},
	url = {https://www.biorxiv.org/content/10.1101/2019.12.17.880153v1},
	doi = {10.1101/2019.12.17.880153},
	language = {en},
	urldate = {2020-01-27},
	journal = {bioRxiv},
	author = {Logiaco, Laureline and Abbott, L. F. and Escola, Sean},
	month = dec,
	year = {2019},
	pages = {2019.12.17.880153}
}

@article{eliasmith_large-scale_2012,
	title = {A {Large}-{Scale} {Model} of the {Functioning} {Brain}},
	volume = {338},
	copyright = {Copyright © 2012, American Association for the Advancement of Science},
	issn = {0036-8075, 1095-9203},
	url = {https://science.sciencemag.org/content/338/6111/1202},
	doi = {10.1126/science.1225266},
	language = {en},
	number = {6111},
	urldate = {2019-04-29},
	journal = {Science},
	author = {Eliasmith, Chris and Stewart, Terrence C. and Choo, Xuan and Bekolay, Trevor and DeWolf, Travis and Tang, Yichuan and Rasmussen, Daniel},
	month = nov,
	year = {2012},
	pmid = {23197532},
	pages = {1202--1205}
}

@book{eliasmith_how_2013,
	title = {How to {Build} a {Brain}: {A} {Neural} {Architecture} for {Biological} {Cognition}},
	isbn = {978-0-19-979454-6},
	shorttitle = {How to {Build} a {Brain}},
	language = {en},
	publisher = {OUP USA},
	author = {Eliasmith, Chris},
	month = jun,
	year = {2013},
	note = {Google-Books-ID: BK0YRJPmuzgC}
}

@article{russo_motor_2018,
	title = {Motor {Cortex} {Embeds} {Muscle}-like {Commands} in an {Untangled} {Population} {Response}},
	volume = {97},
	issn = {0896-6273},
	url = {http://www.sciencedirect.com/science/article/pii/S0896627318300072},
	doi = {10.1016/j.neuron.2018.01.004},
	number = {4},
	journal = {Neuron},
	author = {Russo, Abigail A. and Bittner, Sean R. and Perkins, Sean M. and Seely, Jeffrey S. and London, Brian M. and Lara, Antonio H. and Miri, Andrew and Marshall, Najja J. and Kohn, Adam and Jessell, Thomas M. and Abbott, Laurence F. and Cunningham, John P. and Churchland, Mark M.},
	month = feb,
	year = {2018},
	pages = {953--966.e8}
}

@article{mante_context-dependent_2013,
	title = {Context-dependent computation by recurrent dynamics in prefrontal cortex},
	volume = {503},
	issn = {0028-0836, 1476-4687},
	url = {http://www.nature.com/articles/nature12742},
	doi = {10.1038/nature12742},
	language = {en},
	number = {7474},
	urldate = {2018-04-12},
	journal = {Nature},
	author = {Mante, Valerio and Sussillo, David and Shenoy, Krishna V. and Newsome, William T.},
	month = nov,
	year = {2013},
	pages = {78--84}
}

@article{gao_simplicity_2015,
	series = {Large-{Scale} {Recording} {Technology} (32)},
	title = {On simplicity and complexity in the brave new world of large-scale neuroscience},
	volume = {32},
	issn = {0959-4388},
	url = {http://www.sciencedirect.com/science/article/pii/S0959438815000768},
	doi = {10.1016/j.conb.2015.04.003},
	urldate = {2017-04-06},
	journal = {Current Opinion in Neurobiology},
	author = {Gao, Peiran and Ganguli, Surya},
	month = jun,
	year = {2015},
	pages = {148--155}
}

@article{mastrogiuseppe_linking_2018,
	title = {Linking {Connectivity}, {Dynamics}, and {Computations} in {Low}-{Rank} {Recurrent} {Neural} {Networks}},
	volume = {99},
	issn = {0896-6273},
	url = {http://www.sciencedirect.com/science/article/pii/S0896627318305439},
	doi = {10.1016/j.neuron.2018.07.003},
	number = {3},
	urldate = {2019-01-09},
	journal = {Neuron},
	author = {Mastrogiuseppe, Francesca and Ostojic, Srdjan},
	month = aug,
	year = {2018},
	pages = {609--623.e29}
}

@article{sussillo_generating_2009,
	title = {Generating {Coherent} {Patterns} of {Activity} from {Chaotic} {Neural} {Networks}},
	volume = {63},
	issn = {0896-6273},
	url = {http://www.sciencedirect.com/science/article/pii/S0896627309005479},
	doi = {10.1016/j.neuron.2009.07.018},
	number = {4},
	urldate = {2018-11-08},
	journal = {Neuron},
	author = {Sussillo, David and Abbott, L. F.},
	month = aug,
	year = {2009},
	pages = {544--557}
}

@article{rivkind_local_2017,
	title = {Local {Dynamics} in {Trained} {Recurrent} {Neural} {Networks}},
	volume = {118},
	url = {https://link.aps.org/doi/10.1103/PhysRevLett.118.258101},
	doi = {10.1103/PhysRevLett.118.258101},
	number = {25},
	journal = {Physical Review Letters},
	author = {Rivkind, Alexander and Barak, Omri},
	month = jun,
	year = {2017},
	pages = {258101}
}

@article{thalmeier_learning_2016,
	title = {Learning {Universal} {Computations} with {Spikes}},
	volume = {12},
	issn = {1553-7358},
	url = {http://journals.plos.org/ploscompbiol/article?id=10.1371/journal.pcbi.1004895},
	doi = {10.1371/journal.pcbi.1004895},
	number = {6},
	urldate = {2017-04-27},
	journal = {PLOS Computational Biology},
	author = {Thalmeier, Dominik and Uhlmann, Marvin and Kappen, Hilbert J. and Memmesheimer, Raoul-Martin},
	month = jun,
	year = {2016},
	pages = {e1004895}
}

@article{beer_dynamics_2018,
	title = {Dynamics of dynamics: following the formation of a line attractor},
	shorttitle = {Dynamics of dynamics},
	journal = {arXiv preprint arXiv:1805.09603},
	author = {Beer, Chen and Barak, Omri},
	year = {2018}
}

@article{depasquale_using_2016,
	title = {Using firing-rate dynamics to train recurrent networks of spiking model neurons},
	url = {http://arxiv.org/abs/1601.07620},
	urldate = {2017-01-01},
	journal = {arXiv preprint arXiv:1601.07620},
	author = {DePasquale, Brian and Churchland, Mark M. and Abbott, L. F.},
	year = {2016}
}

@article{boerlin_predictive_2013,
	title = {Predictive {Coding} of {Dynamical} {Variables} in {Balanced} {Spiking} {Networks}},
	volume = {9},
	url = {http://dx.doi.org/10.1371/journal.pcbi.1003258},
	doi = {10.1371/journal.pcbi.1003258},
	number = {11},
	urldate = {2013-11-21},
	journal = {PLoS Comput Biol},
	author = {Boerlin, Martin and Machens, Christian K. and Denève, Sophie},
	month = nov,
	year = {2013},
	pages = {e1003258}
}

@book{eliasmith_neural_2004,
	title = {Neural engineering: {Computation}, representation, and dynamics in neurobiological systems},
	shorttitle = {Neural engineering},
	urldate = {2013-07-28},
	publisher = {MIT Press},
	author = {Eliasmith, Chris and Anderson, C. Charles H.},
	year = {2004}
}

@inproceedings{boerlin_balanced_2012,
	address = {Ohio},
	title = {Balanced spiking networks can implement dynamical systems with predictive coding},
	author = {Boerlin, M and Machens, C. K and Deneve, S},
	year = {2012}
}

@article{sussillo_opening_2013,
	title = {Opening the {Black} {Box}: {Low}-dimensional dynamics in high-dimensional recurrent neural networks},
	volume = {25},
	shorttitle = {Opening the {Black} {Box}},
	url = {http://www.mitpressjournals.org/doi/full/10.1162/NECO_a_00409},
	number = {3},
	journal = {Neural Computation},
	author = {Sussillo, David and Barak, Omri},
	year = {2013},
	pages = {626--649}
}

@article{ben-yishai_theory_1995,
	title = {Theory of orientation tuning in visual cortex},
	volume = {92},
	url = {http://www.pnas.org/content/92/9/3844.short},
	number = {9},
	urldate = {2012-10-03},
	journal = {Proceedings of the National Academy of Sciences},
	author = {Ben-Yishai, R. and Bar-Or, R. L. and Sompolinsky, H.},
	year = {1995},
	pages = {3844--3848}
}

@article{hopfield_neural_1982,
	title = {Neural networks and physical systems with emergent collective computational abilities},
	volume = {79},
	url = {http://www.pnas.org/content/79/8/2554.short},
	number = {8},
	urldate = {2012-09-12},
	journal = {Proceedings of the national academy of sciences},
	author = {Hopfield, J. J.},
	year = {1982},
	pages = {2554}
}

\section{Appendix A -- Nonlinear feature dynamics}

If we consider a desired nonlinear dynamics in the feature space:
\begin{equation}
  \dot{\hat{x}}=-\hat{x}+h(\hat{x})  
\end{equation}

\noindent we replace the definition of $W$ (Equation \ref{eq:identity}) by the following:

\begin{equation}
h(x) = Wf(\phi x)
\end{equation}
obtaining the desired dynamics:

\begin{equation}
\frac{d\hat{x}}{dt} = -\hat{x} + Wf(\phi Wr)
\end{equation}

As above, the temporal derivative of $\hat{x}$ is
\begin{equation}
\frac{d\hat{x}}{dt}=W\frac{dr}{dt}=-\hat{x}+Wf(Jr)
\end{equation}

\noindent thus, by defining 
\begin{equation}
J=\phi W,
\end{equation}
we obtain the desired dynamics. For the linear case, the matrix $A$ will be absorbed into the $W$ matrix, and thus into the final $J$, leading to the same solution.

\section{Appendix B -- Eigenvalues of Jacobian }
The eigenvalues for the Jacobian in Equation \ref{eq:Jacobian} can be derived by observing that the eigenvectors are $\sin \theta$ and $\cos \theta$ truncated between $\pm \theta_C$. We then use $g_1$ from Equation \ref{eq:W_and_g1}:

\begin{eqnarray}
g_{1} &=& \int\frac{d\theta}{2\pi}[\cos\theta-\cos\theta_C]_+\cos\theta \\
 &=& \frac{\theta_C-1/2 \sin2\theta_C}{2\pi}
\end{eqnarray}

To calculate the eigenvalues:

\begin{eqnarray}
\int_{-\theta_C}^{\theta_C}\frac{d\theta}{2\pi}\cos(\theta-\theta')\sin\theta' &=& \frac{\theta_C-1/2 \sin2\theta_C}{2\pi}\sin\theta \\
\int_{-\theta_C}^{\theta_C}\frac{d\theta}{2\pi}\cos(\theta-\theta')\cos\theta' &=& \frac{\theta_C+1/2 \sin2\theta_C}{2\pi}\cos\theta \\
\lambda_{1} &=& 0 \\
\lambda_2 &=& -1 + \frac{\theta_C+1/2 \sin2\theta_C}{\theta_C-1/2 \sin2\theta_C}
\end{eqnarray}

\section{Appendix C -- Adding a bias term}

In the main text we show that NEF recovers the ring model with the constraint of $J_0=0$. From our derivation, we observe that this is due to NEF constraining only the first Fourier mode of the connectivity. We can artificially extend the feature space, in order to allow a non-zero bias term:

\begin{equation}
x=\left(\begin{array}{c}
\cos\psi\\
\sin\psi \\
a
\end{array}\right)
\end{equation}

The tuning curves also need to be artifically extended:

\begin{equation}
\phi=\left(\begin{array}{ccc}
\cos\theta & \sin\theta & -b \end{array}\right)
\end{equation}

Solving the NEF equation indeed leads to a nonzero $J_0=\frac{-ab}{\tilde{g_0}}$, where $\tilde{g}_{0}(\theta_C)=\int\frac{d\theta}{2\pi}[\cos\theta-\cos\theta_C]_+$.

This technical trick is somewhat similar to adding a bias in the Perceptron problem. Note that this ad-hoc solution requires knowledge of the ring model phase diagram, and how it relates to the NEF solutions, and is thus not a general solution.

\section{Appendix D -- Moving bump connectivity}
Below is the derivation of the connectivity for the case of a moving bump. The low-dimensional dynamics are given by:

\begin{equation}
\frac{dx}{dt}=-x+\left(\begin{array}{cc}
1 & v\\
-v & 1
\end{array}\right)x
\end{equation}

which leads to 
\begin{equation}
    \phi A x = \cos (\psi-\theta) + v \sin (\psi-\theta)
\end{equation}

We can write the condition $Ax=f(\phi A x)$ as:

\begin{equation}
(1-i v)e^{i\psi}=\int\frac{d\theta}{2\pi}w(\theta)h(\theta-\psi)
\end{equation}

with

\begin{equation}
    h(z) = f( \cos z + v \sin z )
\end{equation}

Moving to Fourier, the equation becomes
\begin{equation}
    (1-iv)\delta_{k,1} = w_k h_k
\end{equation}

yielding the Fourier coefficient of $w$:
\begin{equation}
    w_k = \delta_{k,1} h_1^{-1} (1-iv)
\end{equation}

The scalar $h_1^{-1}$ can be computed, and will depend on $v$, but we only consider the angular dependence  of $J=\phi A W$ here, arriving at

\begin{equation}
J_{\theta,\theta'} \propto \left[\cos(\theta-\theta')+v\sin(\theta-\theta')\right]
\end{equation}

\section{Appendix E -- Effect of choosing different relevant $x$ values}

\begin{figure*}
		\centering
		\includegraphics[width=0.95\linewidth]{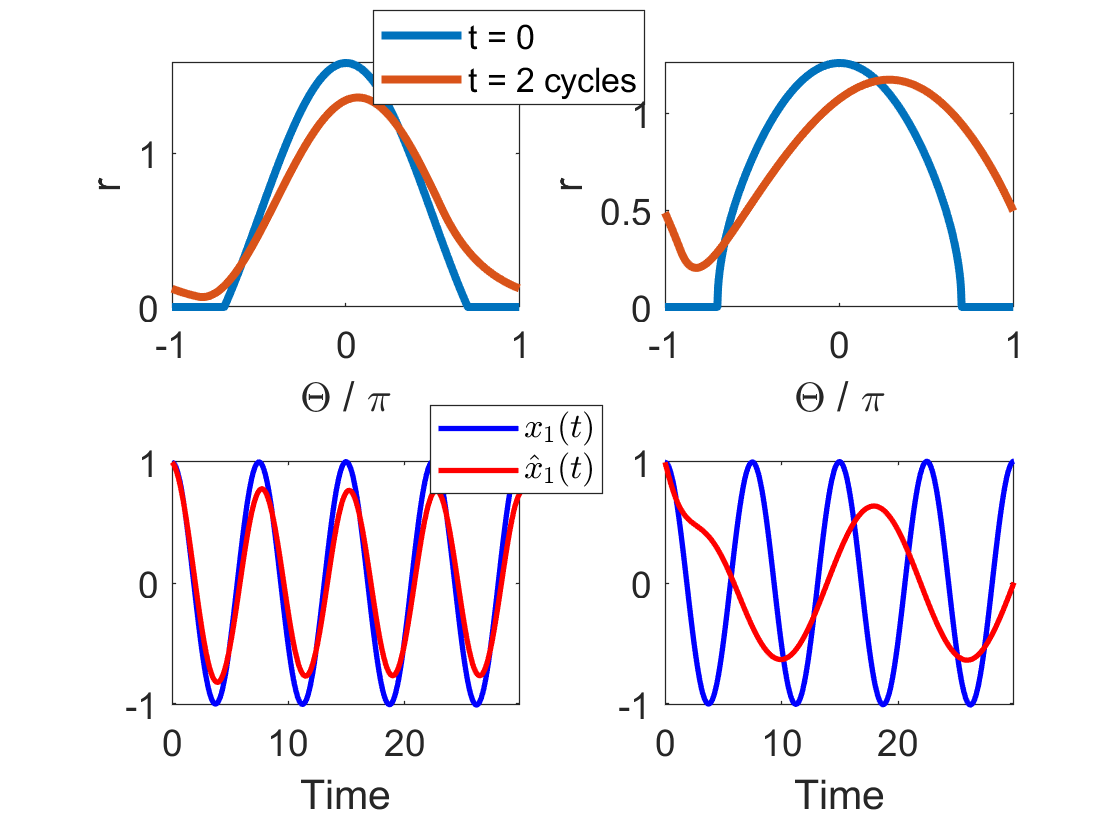}		
		\caption{\textbf{Deviation from desired low-dimensional dynamics.}  When constructing the NEF with the identity of Equation \ref{eq:identity} based on $x$ instead of on $Ax$, the desired low-dimensional dynamics are no longer guaranteed. This is illustrated by a moving bump, similar to Figure \ref{fig:moving}. \textbf{A,C} $f(z) = \left[ z+I_e \right]_+$ \textbf{B,D} $f(z) = \left[ z+I_e \right]_+ ^{0.5}$.}
		\label{fig:moving_nonlin}
	\end{figure*}

In our derivation of the NEF, we implicitly defined the $x$ values of interest (and thus $W$) in equality \ref{eq:identity} as $Ax$. This condition is important, because in general it is not possible to have a consistent mapping for all $x$ values when using linear mappings to approximate nonlinear ones.
For instance, if one used $x$ instead of $Ax$, this would not make any difference for a fixed point dynamics (because $A=I$), and thus the original ring model can still be recovered.
When considering moving bumps, however, this choice results in a different connectivity,  leading to deviations from the desired low-dimensional dynamics. Figure \ref{fig:moving_nonlin} exemplifies this phenomenon for both the threshold-linear function and an additional nonlinearity. In both cases, we see a deviation from the desired low-dimensional manifold. Additionally, in one case (Figure \ref{fig:moving_nonlin}C), both the desired amplitude and the desired frequency are not maintained in the low dimensional feature dynamics.

\end{document}